\documentclass{aastex631}

\begin{document}

% Definitions
\def \planck {\emph{Planck}}
\def \herschel {\emph{Herschel}}
\def \mum {$\rm \mu m$}
\def \lir {$\rm L_{IR}$}
\def \mstar {$\rm M_{*}$}
\def \mdust {$\rm M_{dust}$}
\def \tc {$\rm t_{c}$}
\def \tw {$\rm t_{w}$}
\def \lco {$\rm L^{'}_{CO}$}
\def \lrad {$\rm L_{1.4 GHz}$}
\def \msun{$\rm M_{\odot}$}
\def \lsun{$\rm L_{\odot}$}
\def \rhom{$\rm \rho_{(M_{mol})}$}
\def \maslim {$\rm M_{lim}$}
\def \tcool {$\rm t_{cool}$}
\def \mh{$\rm M_{200}$}
\def \fsf{$f_{\rm SF}$}
\def \fq{$f_{\rm Q}$}
\def \z{$z_{\rm spec}$}

%\title{Unveiling the SMGs properties in SMACS0723 by {\it JWST} {\color{purple}and ALMA} }
\title{Properties of host galaxies of submillimeter sources as revealed by {\it JWST} Early Release Observations in SMACS J0723.3-7327}

\correspondingauthor{Cheng Cheng}
\email{chengcheng@nao.cas.cn}

\author[0000-0003-0202-0534]{Cheng Cheng}
\affiliation{Chinese Academy of Sciences South America Center for Astronomy, National Astronomical Observatories, CAS, Beijing 100101, China}
\affiliation{CAS Key Laboratory of Optical Astronomy, National Astronomical Observatories, Chinese Academy of Sciences, Beijing 100101, China}

\author[0000-0001-7592-7714]{Haojing Yan} %%% yanhaojing@gmail.com 
\affiliation{Department of Physics and Astronomy, University of Missouri,
Columbia, MO 65211}
\author[0000-0001-6511-8745]{Jia-Sheng Huang} %%% jhuang889494@gmail.com
\affiliation{Chinese Academy of Sciences South America Center for Astronomy, National Astronomical Observatories, CAS, Beijing 100101, China} 
\affiliation{Center for Astrophysics, Harvard \& Smithsonian$|$Cambridge, MA 02138, USA}
\author[0000-0001-9262-9997]{Christopher N. A. Willmer} %%% cnawillmer@gmail.com 
\affiliation{Steward Observatory, University of Arizona, 933 N Cherry Ave,
Tucson, AZ, 85721-0009}
\author[0000-0003-3270-6844]{Zhiyuan Ma} %%% zhiyuanma@umass.edu 
\affiliation{Department of Astronomy, University of Massachusetts, Amherst,
USA, 01003}
\author[0000-0002-6642-7483]{Gustavo Orellana-Gonz\'alez} %%% gustavo.orellana.gonzalez@gmail.com
\affiliation{Instituto de Investigaci\'on Interdisciplinaria, Universidad de Talca, Avenida Lircay, Talca, Chile}

%% Note that the \and command from previous versions of AASTeX is now
%% depreciated in this version as it is no longer necessary. AASTeX 
%% automatically takes care of all commas and "and"s between authors names.

%% AASTeX 6.31 has the new \collaboration and \nocollaboration commands to
%% provide the collaboration status of a group of authors. These commands 
%% can be used either before or after the list of corresponding authors. The
%% argument for \collaboration is the collaboration identifier. Authors are
%% encouraged to surround collaboration identifiers with ()s. The 
%% \nocollaboration command takes no argument and exists to indicate that
%% the nearby authors are not part of surrounding collaborations.

%% Mark off the abstract in the ``abstract'' environment. 
\begin{abstract}
Using the 0.9--4.4~$\mu$m imaging data from the James Webb Space Telescope 
({\it JWST}) Early Release Observation in the SMACS J0723.3-7327 galaxy cluster
field, we discuss the properties of three submillimeter galaxies (SMGs)
detected by the Atacama Large Millimeter Array. These sources are magnified by 1.4--2.1$\times$due to gravitational
lensing. This is the first time that SMG host galaxies are resolved in the
rest-frame near-infrared (NIR). 
One source was previous undetected by HST, while the remaining two are disk galaxies with S\'ersic indices of $\sim 0.9$ and star formation rates on or just below the star formation ``main sequence".
Their submillimeter emission originates from the inner parts
of the hosts, suggesting that their dust contents are concentrated towards the
center. The host half-light radii measured in the rest-frame NIR are 
$\sim$1.5$\times$ smaller than those measured in the rest-frame optical, consistent with a concentrated dust distribution. 
The more severe extinction that optical light suffers towards the center makes it seemingly less concentrated. 
Therefore, we expect that the optically-based determination of the stellar mass
distribution within host galaxies could still be severely biased by dust.
Interestingly, these two disk galaxies are dramatically different in their
outer regions, with one being star forming and the other being quiescent. 
Upcoming {\it JWST} observations of statistically significant samples of SMGs will allow us to understand the correlation between the dusty star forming regions and their hosts.
\end{abstract}
%% Keywords should appear after the \end{abstract} command. 
%% The AAS Journals now uses Unified Astronomy Thesaurus concepts:
%% https://astrothesaurus.org
%% You will be asked to selected these concepts during the submission process
%% but this old "keyword" functionality is maintained in case authors want
%% to include these concepts in their preprints.

%% From the front matter, we move on to the body of the paper.
%% Sections are demarcated by \section and \subsection, respectively.
%% Observe the use of the LaTeX \label
%% command after the \subsection to give a symbolic KEY to the
%% subsection for cross-referencing in a \ref command.
%% You can use LaTeX's \ref and \label commands to keep track of
%% cross-references to sections, equations, tables, and figures.
%% That way, if you change the order of any elements, LaTeX will
%% automatically renumber them.
%%
%% We recommend that authors also use the natbib \citep
%% and \citet commands to identify citations.  The citations are
%% tied to the reference list via symbolic KEYs. The KEY corresponds
%% to the KEY in the \bibitem in the reference list below. 

\section{Introduction} \label{sec:intro}
The Sub-millimeter galaxies (SMGs, \citealt{sma98,bla02}) are a population of high redshift, massive \citep{dud20}, gas-rich \citep{tac06,rie10}, and high star formation rate \citep[SFR,][]{bar14} galaxies, with considerable emission in the sub-millimeter (submm, \citealt{cas14}), as a consequence of the high dust grain attenuation. Meanwhile, the SMGs are also associated with active galactic nuclei (AGNs,\citealt{ued18}).

In spite of the intense study of the SMGs in the past two decades, these galaxies are still poorly understood because of their dusty nature. In addition, due to their high redshifts and dust extinction, the observed photometry in the optical bands data are probing SMGs in the rest-frame UV, which also limits the ability to constrain  the spectral energy distribution (SED) and derived properties such as  the photometric redshift (photo-z) and stellar mass (\mstar). Previous SMG morphology studies found that SMGs are mainly disky \citep{2014ApJ...782...68T}. Because the \emph{Hubble Space Telescope} ({\it HST}) imaging only probes SMGs out to the rest-frame $B$ to $V$ bands, the compact starburst in the central dusty region \citep{2019MNRAS.490.4956G, 2017ApJ...850...83F} may also lead to a flat morphology. The \emph{Spitzer}/IRAC images have been fundamental to study the stellar properties of SMGs \citep[e.g.,][]{2021MNRAS.505.1509S}, but do have a limitation because of the image resolution of $\sim$  2$"$, so that SMGs are frequently blended in the deep IRAC images, preventing us from accurately measuring the SEDs and spatial properties.

The recent launch and the Early Release Observations \citep[ERO,][]{2022arXiv220713067P} of the \emph{James Webb Space Telescope} ({\it JWST}) provide a new window to explore SMGs in the unrivaled depth and resolution in the Mid-Infrared (MIR, $\lambda_{\rm rest}$ $>$ 2.5 $\mu$m). The ERO of {\it JWST} contains observations of the Reionization Lensing Cluster Survey \citep[RELICS,][]{coe19} cluster SMACS J0723.3-7327 (hereafter SMACS0723) at $z=0.4$. Prior {\it HST} observations of SMACS0723 used the {\it HST}/ACS camera in F435W, F606W, F814W band, and the WFC3 camera in the F105W, F125W, F140W, F160W. All the {\it HST} images reach to about 26.5 AB mag. 
SMACS0723 is also mapped by the Spitzer Reionization Lensing Cluster Survey (SRELICS{\footnote{https://www.ipac.caltech.edu/doi/irsa/10.26131/IRSA426}}, PI: Brada{\v{c}}), 
and the ALMA Lensing Cluster Survey \citep[ALCS, PI: Kotaro Kohno;][]{2022ApJ...932...77S, 2022arXiv220707125K} in band 6 (260 GHz, 1.15mm). Gravitational lensing by the galaxy cluster boosts the background galaxy flux, facilitating the detection of high-z galaxies. Therefore, the combination of the archival data and the MIR data from {\it JWST} ERO observations of SMACS0723 provide us with a very unique chance to study SMGs behind SMACS0723. We adopt $H_0 = 70$~km~s$^{-1}$~Mpc$^{-1}$, $\Omega_M=0.3$, $\Omega_\Lambda=0.7$, and AB magnitudes \citep{1983ApJ...266..713O} throughout.

\section{SMG sample and multi-wavelength data sets}

\subsection{SMGs from ALMA band 6 image}
we use archival data for the ALMA Band 6 ALCS observations of SMACS0723 \citep{2022ApJ...932...77S, 2022arXiv220707125K}. The data were reduced and calibrated using the default {\sc ScriptForPI.py} script in the Common Astronomy Software Applications \citep[CASA, ][]{2007ASPC..376..127M}. The continuum map is mosaiced by {\sc tclean} with the parameters {\sc {gridder=`mosaic', weighting=``briggs", robust=2.0}}. The final image mosaic reaches an rms of about 0.066~mJy with a beam size of 1\farcs01$\times$0\farcs77, $\rm PA = 22$\arcdeg. 

To perform the target identification and photometry we used
Source Extractor \citep[SExtractor,][]{1996A&AS..117..393B}) with a detection threshold of 4 $\sigma$ above the background noise on the primary beam uncorrected map, which leads to 6 targets located within the NIRCam field of view. To validate the target selection, we check the negative image, where one target is found; this translates into a reliability of $(N_{\rm positive} - N_{\rm negative})/N_{\rm positive} = 83.3\%$. To further validate the results, we use the ALMA coordinates to find NIRCam counterparts within a $0\arcsec.5$ search radius. The 3 objects that have clear NIRCam counterparts are the subject of this paper. We include 10\% of the flux as the flux error to account for the calibration uncertainty \citep{2014Msngr.155...19F} .

\begin{deluxetable*}{cccccccccc}
\tabletypesize{\footnotesize}%footnotesize}%\scriptsize}%%%\small}
\tablewidth{0pt}
\tablecaption{SMG catalog in SMACS0723.\tablenotemark{\rm *} }
\label{t:targets}
\tablehead{
Name    & RA      & Dec      & redshift    &  $S_{260 \rm GHz}$ & log(\mstar) & log(SFR)  & Half-light radius$_{\rm 1.1mm}$ & Half-light radius$_{\rm F444W}$ & magnification factor\\ %\tablenotemark{a} \\ 
        &(J2000)  &  (J2000) &              & (mJy)                 & (\msun)      & (\msun/yr)   & (arcsec) & (arcsec)  & $\mu$
}
\startdata
ID1 & 07:23:15.1 & -73:27:46.2 & 0.89\tablenotemark{\rm **} & 0.58$\pm$0.06 & --  & --  & --  & 0.15$\pm$0.10 & 1.36 \\
ID2 & 07:23:03.9 & -73:27:06.1 & 2.463 & 1.28$\pm$0.13                   & 11.81$\pm$0.11 & 2.10$\pm$0.55 & 0.44 $\pm$ 0.25 & 0.80$\pm$0.10 & 2.08  \\
ID3 & 07:23:25.1 & -73:27:38.9 & 1.05  & 0.42$\pm$0.04                   & 10.43$\pm$0.07 & 1.37$\pm$0.11 & 0.28 $\pm$ 0.22 & 0.49$\pm$0.10 & 1.69 \\
\enddata
\tablenotetext{*}{The flux, \mstar, SFR and half-light radii values in this table are not corrected by the lensing magnification factor. The typical uncertainty of the magnification factor is 20\% \citep{2022arXiv220707102P, 2022arXiv220707101M}.}
\tablenotetext{**}{The photo-z and physical parameters for ID1 may not be reliable. This target is more likely to be an SMG at redshfit above 4, see Sec. \ref{disid1}.}
\end{deluxetable*}

\begin{figure}
    \centering
    \includegraphics[width = 0.95\textwidth]{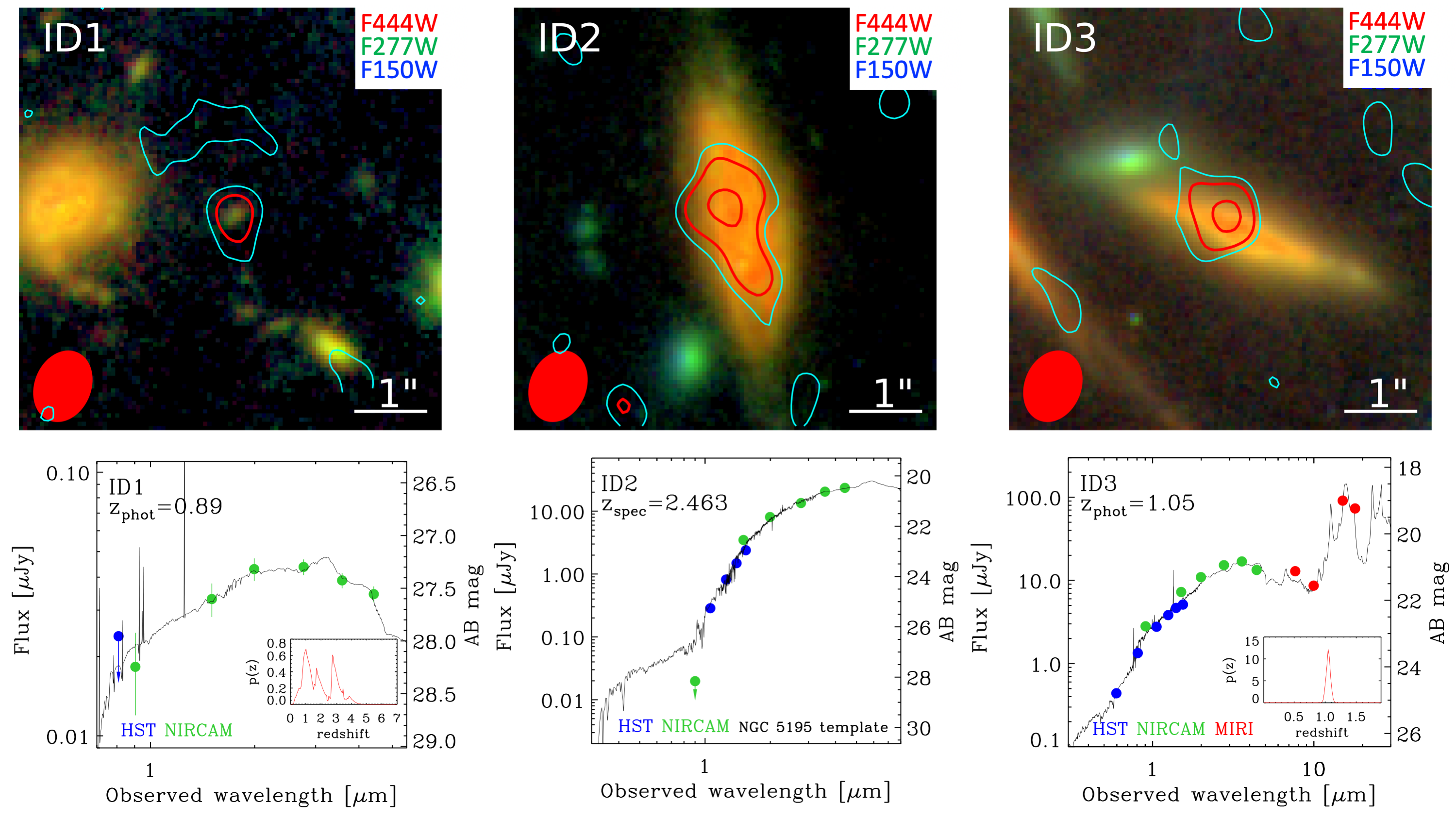}
    \caption{
    {\bf Upper panels:} 
    Color composite images of the SMGs in this work combining the F150W (blue), F277W (green) and F444W (red) bands. The ALMA continuum is represented by contours in 2$\sigma$ (cyan) and $3\sigma$, $5\sigma$ (red) of the ALMA continuum map noise. The ALMA beams are shown as filled red ellipses of each panel. {\bf Bottom panels:} Best fit SEDs of our sample. We show the photometry from {\it HST} (blue), {\it JWST}/NIRCam (green) and {\it JWST}/MIRI (red). Since the spec-z of the target ID2 is 2.464, we fit the SED with local galaxy templates \citep{2014ApJS..212...18B}, and find the best-fitted template is the one of NGC5195. ID1 and ID3 have not been spectroscopically confirmed. Therefore we estimate the photo-z with EAZY, showing the templates with highest probability, and the p(z) v.s. redshift in the subplot panel. The MIRI photometry matches well with the PAH feature that predicted by the best-fitted template. ID1 is the faintest target in our sample and the p(z) shows two peaks at $z$ $\sim$ 0.9 and 2.9; we further discuss the ID1 photo-z in Sec. \ref{disid1}.}
    \label{stamp}
\end{figure}

\begin{figure}
    \centering
    \includegraphics[width = 0.6\textwidth]{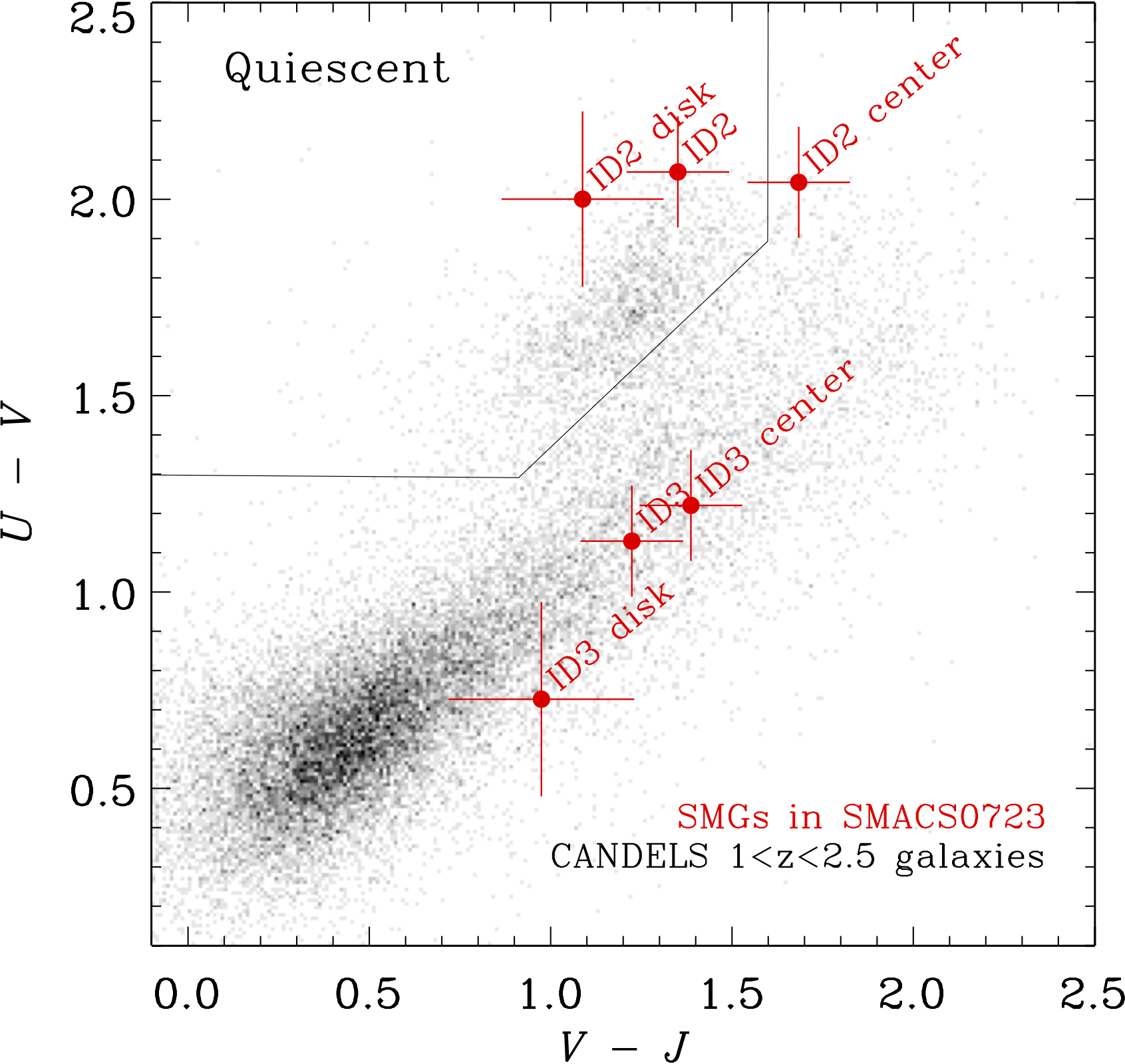}
    \caption{The $V-J$ v.s. $U-V$ diagram of our sample. We show the CANDELS \citep{2011ApJS..197...35G,  2011ApJS..197...36K} sample at redshifts $1<z<2.5$ as a comparison. The solid lines are adopted from \citet{2018ApJ...858..100F} to separate quiescent from star forming galaxies. 
    }
    \label{UVJ}
\end{figure}

\subsection{{\it JWST} data}

The {\it JWST} ERO contains NIRCam images in F090W, F150W, F200W in the short wavelength channel (SW) and F277W, F356W and F444W in the long wavelength channel (LW). The public release includes the processed ``Stage 3'' images. However, the astrometry with these images is not accurate enough for our study. Therefore, we re-processed the data based on the ``Stage 2'' products. A number of issues with the current {\it JWST} data reduction pipeline were fixed, including the pedestal background correction in the SW bands. The final mosaics have the scale of 0\arcsec.06~pix$^{-1}$, and are aligned to the RELICS {\it HST} images in pixel coordinates. We also corrected the zero-points of NIRCam filters using the STScI calibration CRDS file $jwst\_0942.pmap$ which uses the photometric calibration presented in \citet{2022arXiv220705632R}. The astrometry is tied to GAIA DR2 and is accurate to $\sim$30 mas. We show the NIRCam image cutouts and ALMA contours in Fig. \ref{stamp}. This is the first time we resolve the SMG morphology in the rest-frame near-infrared bands; in our sample there are two disk galaxies (ID2 and ID3) and one faint small galaxy (ID1 with F200W magnitude $\sim$ 27).

SMACS0723 is also observed by {\it JWST}/MIRI in the F770W, F1000W, F1500W, and F1800W bands. Of the three SMGs, only ID3 has a MIRI detection; ID1 is undetected while ID2 is not covered by MIRI. Our new results confirm the photo-z measured by \citet[][$z_{\rm phot}=0.9820$]{coe19} and the MIRI observations show good agreement with the presence of polycyclic aromatic hydrocarbons (PAH) commonly found in dusty star-forming galaxies.

\subsection{Lens model}

Recently, \citet{2022arXiv220705007G} published a lens model for SMASC0723, based on images from the {\it HST} and redshifts measured from the Multi Unit Spectroscopic Explorer \citep[MUSE,][]{2010SPIE.7735E..08B} using the \texttt{Light-Traces-Mass}  (LTM) approach \citep{2015ApJ...801...44Z}. The NIRCam imaging of SMACS0723 revealed new multiply imaged galaxies from which updated parametric lensing models were built \citep{2022arXiv220707102P, 2022arXiv220707101M, 2022arXiv220707567C}. In this work, we adopt the average from the LTM model of 
\citet[][]{2022arXiv220705007G} and the parametric model of \citet[][]{2022arXiv220707102P}.

\section{properties of the SMG in this work}

\subsection{SMG SED fitting}

We used SExtractor in dual-image mode to measure the photometry for the entire cluster field, using F444W as the detection image\footnote{We did not use the PSF matched photometry because the complex kernel may introduce more noise for the three SMGs. We use the {\sc mag\_ ISO} to ensure a more reliable color measurements for the SED fitting, as well as including most of the flux.}. As mentioned in Section 2.2, ID3 is the only target in our sample with both {\it HST} and {\it JWST}/MIRI detections. 
Only ID2 has a spectroscopic redshift \citep[z=2.463,][]{2022arXiv220708778C}, so we derive the photometric redshifts of ID1 and ID3 using EAZY \citep{Bra08} with the default templates and the local galaxy templates \citep{2014ApJS..212...18B}, which include the PAH features to match the MIRI data of ID3. 
We include 0.05 mag uncertainty to alleviate potential impact of yet uncertain instrument calibrations.

The EAZY photo-z and p(z) curves are shown in Fig. \ref{stamp}. For ID2, we fit the SED by the local galaxy templates with spec-z, and find the NGC 5195 template has the minimal $\chi^2$. NGC 5195 is the dusty minor merger galaxy in the M51 system, implying that ID2's SED is consistent with the dusty emission, though the model predicted F090W is above the observed F090W flux 5$\sigma$ upper limit.

We use the Multi-wavelength Analysis of Galaxy Physical Properties (MAGPHYS) code \citep{Cun08} to fit the SEDs including the {\it HST}, {\it JWST} and ALMA flux simultaneously, and measure the physical properties such as the $M_*$ and SFR (Table 1). The fitting results are shown in the appendix  Fig. \ref{magphys}. The MAGPHYS fit for ID1 shows a level of dust emission higher than the model prediction, suggesting that the EAZY photo-z may not be reliable (see Sect. \ref{disid1}), and because of this, we do not list its M$_*$ and SFR in Table 1. In Fig. 1 we can see the dust emission contours of ID2 and ID3 are more concentrated towards the galaxy center, while the MAGPHYS dust extinction is assumed as a galaxy-wide average. 
Given the limited information from the ALMA maps, we adopt the SFR from MAGPHYS. We compare the specific SFR of the main sequence galaxies  \citep[$s{\rm SFR}(M_*, z)$, ][]{2015ApJ...800...20G} and the target ID2, ID3. The ratio between the sSFR of our sample and the main sequence galaxies with their stellar mass and redshift are $s{\rm SFR}_{\rm ID2} / s{\rm SFR_{\rm main\ sequence}} = 0.23 \pm 0.30$, and $s{\rm SFR}_{\rm ID3} / s{\rm SFR_{\rm main\  sequence}} = 0.88\pm0.29$, implying that ID2 is more likely to be a quiescent galaxy candidate, and ID3 is a main sequence galaxy at redshift 1.

To characterize the stellar populations, we derive the {\it U, V, J} rest-frame magnitudes, and show the $V-J$ v.s. $U-V$ of ID2 and ID3 in Fig. \ref{UVJ}. To further reveal the spatial color information resolved by {\it JWST}, we measure the photometry inside the central 0\arcsec.6 ($\sim 5$ kpc at redshift 2) to get the central region flux, and estimate the outer disk flux using the measurements beyond 0\arcsec.6. 
These colors show that the central regions of ID2 and ID3 are dusty. 
Meanwhile, the disk region of ID3 is more like a star forming galaxy, while the disk region of ID2 may have been quenched.

\begin{figure}
    \centering
    \includegraphics[width = 0.6\textwidth]{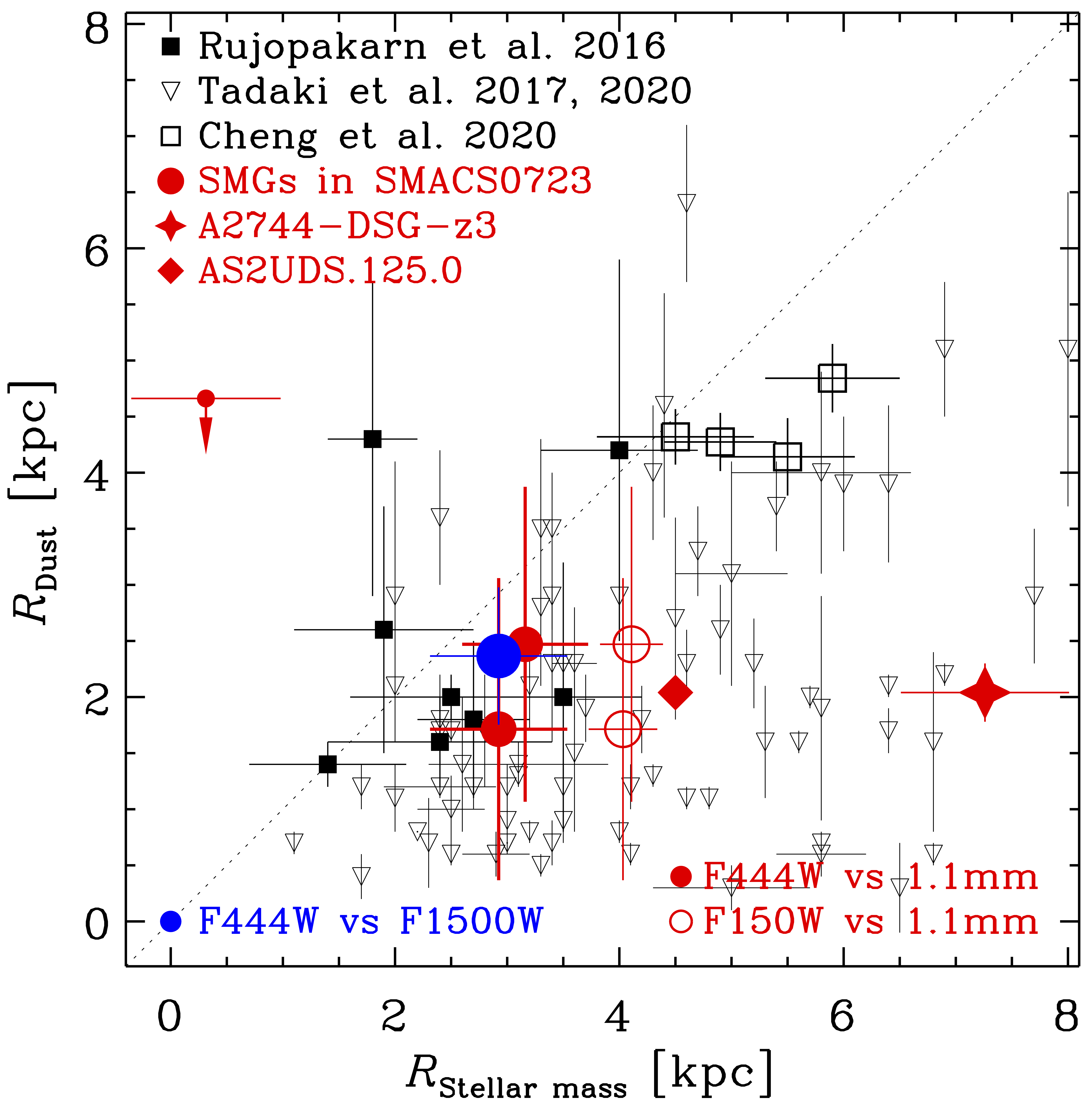}
    \caption{The half-light radius from the NIRCam F444W image (red points) compared to the dust size from ALMA. We also show the F150W half-light radii (red circles), which are $\sim$ 1.5 times larger than the F444W radii, and the SMG size measured 
    from the {\it JWST}/NIRCam images \citep[red diamond and star points,][]{2022arXiv220805296C, 2022arXiv220808473W}, or 
    the size measured from {\it HST} F140W or F160W images \citep[black points, which is roughly considered as the stellar mass distribution radius,][]{2016ApJ...833...12R, 2017ApJ...834..135T, 2020ApJ...901...74T, 2020MNRAS.499.5241C}.
    The size difference between the F150W and F444W bands may be caused by the dust extinction, and implies that SMGs stellar distribution may be smaller than the size revealed by {\it HST} images. The blue circle is the F444W vs F1500W size for ID3, 
    which measures the extents of regions dominated by the stellar component and the PAH emission at $z$$\sim$1.
    }
    \label{size}
\end{figure}

\subsection{Morphology of this SMG sample}

We also measure the ALMA image sizes of the SMGs using {\sc uvmultifit} \citep{2014A&A...563A.136M}, which fits a Gaussian model in {\it uv} space. The deconvolved half-light radius is $\sim$ 2 kpc, which is consistent with the typical SMG dust size, and implying a compact star-forming region at the galaxy center. To compare the stellar mass distribution, we measure the half-light radius of the SMGs from the F444W image, which corresponds to the rest-frame near infrared band, and close to the stellar morphology of the SMGs. We correct for the PSF broadening using $\sigma_{\rm target}^2 - \sigma_{\rm PSF}^2$ with the PSF generated by $WebbPSF$\footnote{\url{https://github.com/mperrin/webbpsf}}. Because previous SMG size measurements are mainly based on {\it HST} 160W, we compare the {\it HST} half-light radius measurements with those of {\it JWST} taken in the F150W band. Finally, we correct the galaxy size by dividing the square root of the magnification factor.

The dust size of our sample is shown in Fig. \ref{size}. Because dust attenuation affects primarily the central parts of galaxies, its presence will bias the half-light radius estimation, the half-light radius increasing as the attenuation in the galaxy center becomes larger. We can see the SMGs in this work also have a concentrated dust distribution, and a larger stellar mass size. Moreover, the F150W half-light radius would be 1.5 times larger than the F444W , thus the stellar distribution size is still large, but may not be as extended as we measured from the {\it HST} images. 

We also measure the half-light radius of ID3 in the F1500W band, which covers the PAH feature at redshift 1. Our result shows that the dust size is marginally more extended than the 1.1mm band size, implying the warm dust from PAH may be more extended than the cold dust detected by ALMA. Higher resolution ALMA observations are still needed to explore the spatial relation between the PAH and 1.1mm emission.

We model the galaxy morphology of ID2 and ID3 in F444W using GALFIT \citep{2002AJ....124..266P} and find a Sersic index $\sim$ 0.9 (Fig. \ref{galfit}). The disky structure of ID2 and ID3 indicates that the compact FIR emission in the galaxy center cannot be caused by major mergers \citep{1988ApJ...325...74S}. 
On the other hand, although both ID2 and ID3 show a disky morphology, 
the disks themselves display different colors in the {\it UVJ} diagram,
which may imply different evolutionary paths for high-z disk galaxies.

\begin{figure}
    \centering
    \includegraphics[width = 0.9\textwidth]{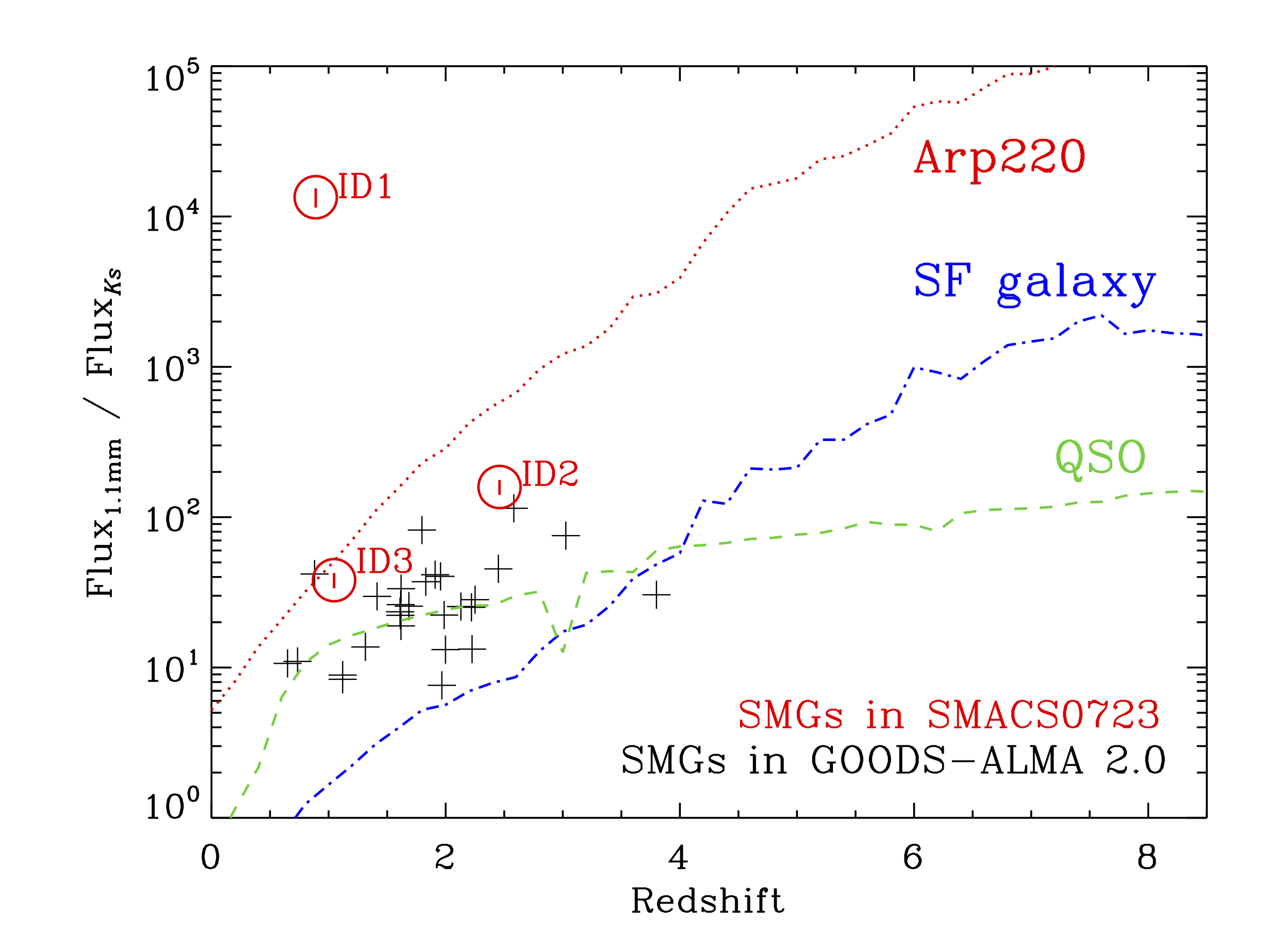}
    \caption{The evolution of the 1.1 mm to $Ks$ observed
    flux ratio as a function of redshift. We show the SMGs from \citet[ALMA-GOODS 2.0]{2022A&A...658A..43G}. The expected trends for a starburst galaxy (Arp220, dotted red line), a typical star-forming galaxy (dot-dashed blue line) and a QSO (dashed green line). Both ID2 and ID3 follow the the trend of Arp220 well, but ID1 shows a flux ratio that is too high, suggesting a high-z nature of this target. 
    }
    \label{submmk}
\end{figure}

\section{Discussion}

\subsection{Offset between submm and NIRCam}
\label{offset}

The offset between the ALMA and optical image positions of the SMGs reflects the displacement of the rest frame UV and FIR, which may be the clumpiness of the star formation region or heavy obscuration by  dust \citep[e.g., ][]{2018ApJ...865..106C}. If the dusty starbursts of the SMGs are in the gravitational center of galaxies, we can expect the stellar mass centers to be better aligned with the dust emission. However, we find that the ID2 and ID3 centers in the F444W image are also offset from ALMA by $\sim$ 0\arcsec.3 ($\sim$2.5 kpc at $z=1.5$). The ALMA pointing accuracy\footnote{https://help.almascience.org/kb/articles/what-is-the-absolute-astrometric-accuracy-of-alma}  is  \ $P_{\rm acc} = FWHM/SNR/0.9 \simeq 0\arcsec.1$. If we include the ALMA system offset by $0\arcsec.06$ \citep{2022ApJ...929...40L} and the NIRCam WCS accuracy $\sim$ $0\arcsec.05$, there is still  $\sim0\arcsec.1$ ($\sim$  0.8 kpc at redshift 2) offset between the 1.1mm and F444W emission
(our astrometry is accurate to ${\rm rms} = 0\arcsec.03$ comparing with GAIA DR2).
. Minor mergers may explain this offset, as well as the dusty feature in the disky galaxies. Higher resolution ALMA observations are still needed to address the origin of the dusty regions.

\subsection{Photo-z of ID1: high-z SMG?}
\label{disid1}

The detection of ID1 by NIRCam demonstrates the sensitivity of the {\it JWST}. The photo-z of 0.9 is mainly driven by the peak at F277W and the drop in the F356W and F444W bands, constraining the template peak at rest-frame 1.6$\mu$m. However, this target is not detected in {\it HST} images, and the large error bar of the flux hardly constrains the template fitting. In Fig. \ref{submmk}, we show the observed 1.1mm and the observed $Ks$ band flux ratio vs redshift for the SMGs from the ALMA-GOODS 2.0 survey \citep[ALMA-GOODS 2.0]{2022A&A...658A..43G} and the templates for SMGs (Arp 220); a star-forming galaxy and a QSO template from \citet{2007ApJ...663...81P}. This figure directly shows the possible redshift range based on the observed flux ratio. We can see the 1.1mm/$Ks$ flux of ID1 is too high at redshift 0.9, and should be at $z>4$, which is also consistent with the red color at F090W - F200. 
On the other hand, visual examination of JWST images has revealed that ID1 is located along the direction of a bright star spike. This in return might contaminate the photometry, and result in erroneous interpretation of the redshift and physical parameters.

\section{Summary}
We present the first study of ALMA detected SMG host galaxies based on the
rest-frame NIR data from {\it JWST}. While our limited sample contains only three 
objects, we have obtained some ``first look'' results of the SMGs.
We find one object previously undetected by {\it HST}, and there is a hint
from its mm-to-IR flux ratio that it could be at $z\gtrsim 4$. The other two
objects are disky galaxies, one of which seems to have spiral arms. This
suggests that their star formation activity is taking place in the secular 
mode instead of being triggered by major mergers. 
This conclusion is corroborated by the SED analysis which shows these galaxies as being on or below the star-formation main sequence. The high resolution
NIRCam images also show that their dust emission regions are concentrated to
the central regions of the hosts. Consequently, their rest-frame NIR images 
are more compact than in the rest-frame optical, as NIR light suffers less dust 
extinction. This reveals a potential bias in the optical-based studies of the stellar mass distribution within galaxies. Lastly, we show that these two galaxies are dramatically different beyond the dust emission regions that one has star forming disk while the other has a quiescent disk. However, further studies of the NIR properties of SMG hosts in statistically significant samples are still necessary to reveal the nature of these objects.

%% IMPORTANT! The old "\acknowledgment" command has be depreciated. It was
%% not robust enough to handle our new dual anonymous review requirements and
%% thus been replaced with the acknowledgment environment. If you try to 
%% compile with \acknowledgment you will get an error print to the screen
%% and in the compiled pdf.
%% 
%% Also note that the akcnowlodgment environment does not support long amounts of text. If you have a lot of people and institutions to acknowledge, do not use this command. Instead, create a new \section{Acknowledgments}.

\begin{acknowledgments}
We thank the referee for the kind and constructive comments that helped us to improve the manuscript. C.C. thanks Fengwu Sun for helpful discussions. We are grateful to the {\it JWST} ERO teams for doing the observations and for the prompt data release. This work is supported by the National Natural Science Foundation of China, No. 11803044, 11933003, 12173045. This work is sponsored (in part) by the Chinese Academy of Sciences (CAS), through a grant to the CAS South America Center for Astronomy (CASSACA). We acknowledge the science research grants from the China Manned Space Project with NO. CMS-CSST-2021-A05. CNAW acknowledges support from the NIRCam Development Contract NAS5-02105 from NASA Goddard Space Flight Center to the University of Arizona.

Some/all of the data presented in this paper were obtained from the Mikulski Archive for Space Telescopes (MAST) at the Space Telescope Science Institute. The specific observations analyzed can be accessed via \dataset[https://doi.org/10.17909/4wr8-hh69]{https://doi.org/10.17909/4wr8-hh69}. STScI is operated by the Association of Universities for Research in Astronomy, Inc., under NASA contract NAS5–26555. Support to MAST for these data is provided by the NASA Office of Space Science via grant NAG5–7584 and by other grants and contracts.

This paper makes use of the following ALMA data: ADS/JAO.ALMA\#2018.1.00035.L. ALMA is a partnership of ESO (representing its member states), NSF (USA) and NINS (Japan), together with NRC (Canada), MOST and ASIAA (Taiwan), and KASI (Republic of Korea), in cooperation with the Republic of Chile. The Joint ALMA Observatory is operated by ESO, AUI/NRAO and NAOJ. The National Radio Astronomy Observatory is a facility of the National Science Foundation operated under cooperative agreement by Associated Universities, Inc. The ALMA data reduction and other data services of this work are fully or partially supported by China-Chile Astronomical Data Center (CCADC), which is affiliated to Chinese Academy of Sciences South America Center for Astronomy (CASSACA). 

\end{acknowledgments}

\appendix

We show the MAGPHYS SED fitting results in Fig. \ref{magphys} and the GALFIT fitting results in Fig. \ref{galfit}.

\setcounter{figure}{0}
\renewcommand{\thefigure}{A\arabic{figure}}

\begin{figure}
   \centering
    \includegraphics[width = 0.8\textwidth]{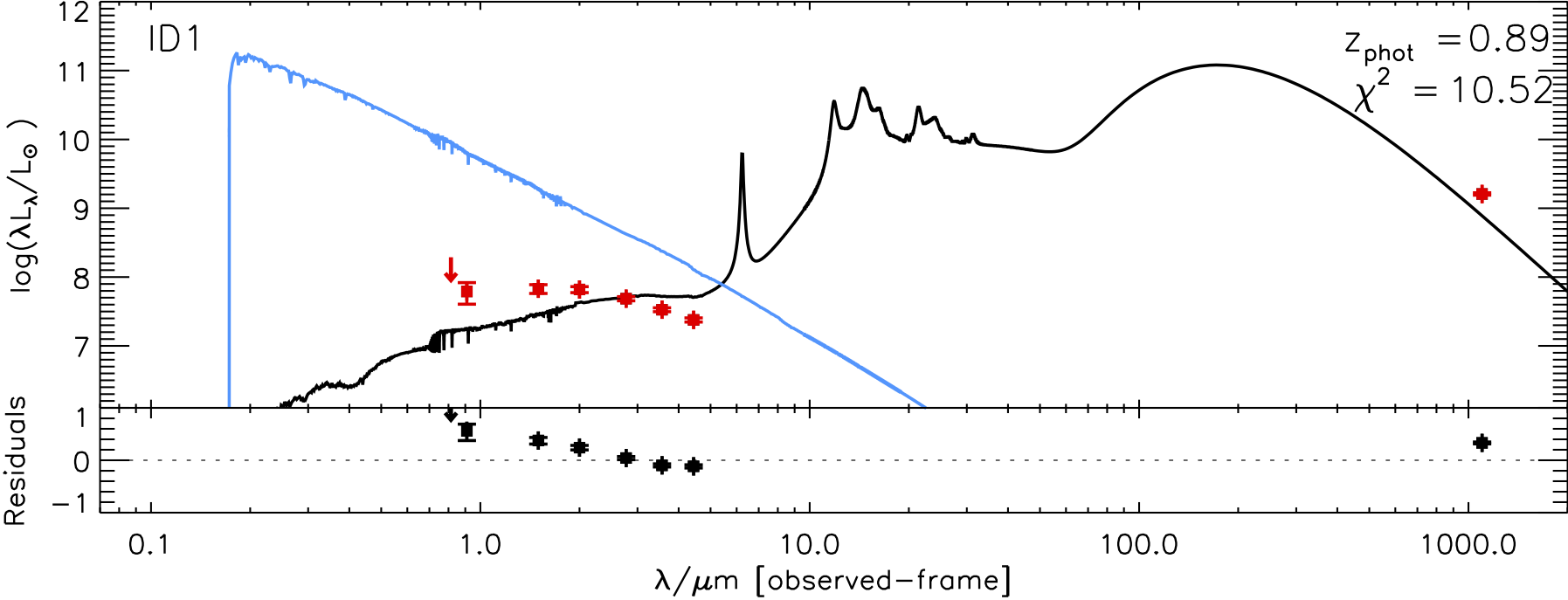}
    \includegraphics[width = 0.8\textwidth]{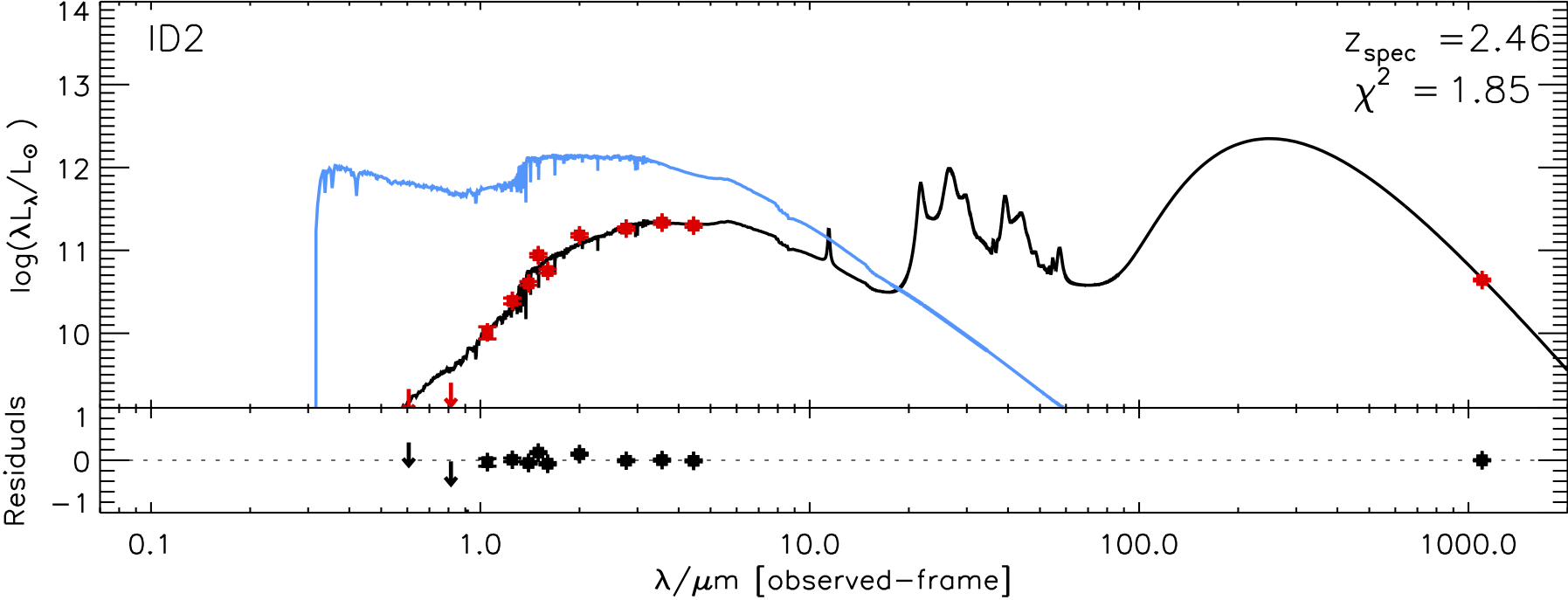}
    \includegraphics[width = 0.8\textwidth]{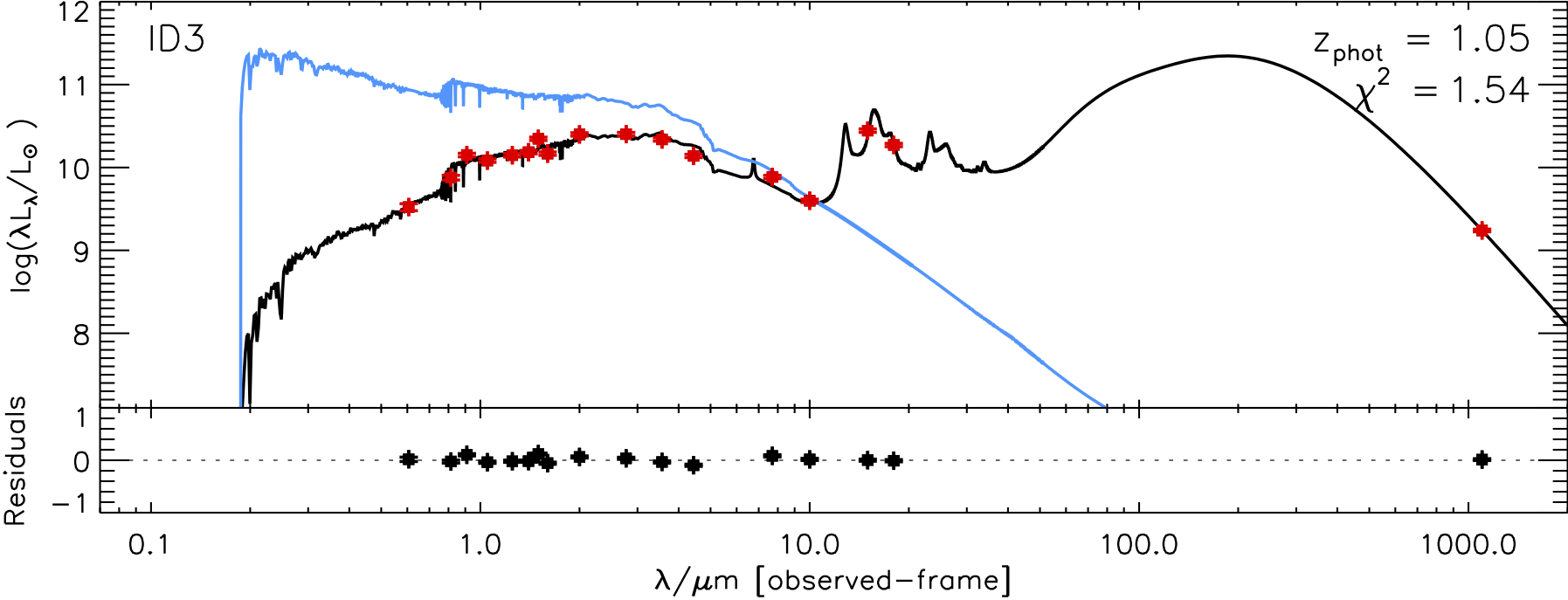}
   \caption{SED fitting results given by MAGPHYS with the observed data (red), best fitted model (black) and the stellar model (blue) before dust extinction, and the residual in each bottom panel.}
   \label{magphys}
\end{figure}

\begin{figure}

    \centering
    \includegraphics[width = 0.8\textwidth]{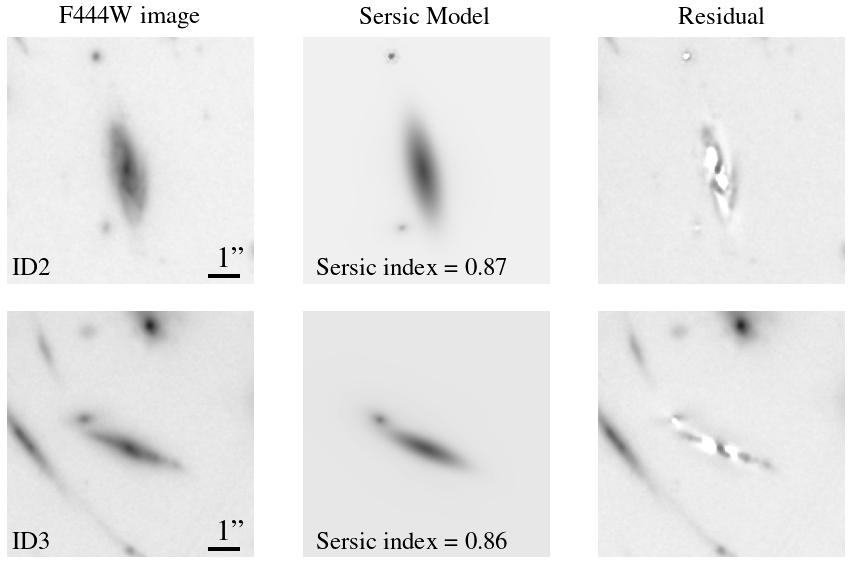}
    \caption{The NIRCam F444W image (left panel), best fitted by a Sersic model (central panel) and the fit residual (right panel) of ID2 and ID3.} % The clear spiral arms in the residual of ID2 is similar to the other spiral galaxies found at redshift 2. \citep{2012Natur.487..338L}. Since the spiral arms may have been formed by the high gas abundance or as a result of a minor merger, dynamical observations would help to understand the origin of the spirals. }
    \label{galfit}
\end{figure}

%% To help institutions obtain information on the effectiveness of their 
%% telescopes the AAS Journals has created a group of keywords for telescope 
%% facilities.
%
%% Following the acknowledgments section, use the following syntax and the
%% \facility{} or \facilities{} macros to list the keywords of facilities used 
%% in the research for the paper.  Each keyword is check against the master 
%% list during copy editing.  Individual instruments can be provided in 
%% parentheses, after the keyword, but they are not verified.

\vspace{5mm}
\facilities{{\it HST}(ACS, WFC3-IR), {\it JWST} (NIRCAM, MIRI), ALMA: band6}

%% Similar to \facility{}, there is the optional \software command to allow 
%% authors a place to specify which programs were used during the creation of 
%% the manuscript. Authors should list each code and include either a
%% citation or url to the code inside ()s when available.

\software{astropy \citep{2013A&A...558A..33A,2018AJ....156..123A},  
          Source Extractor \citep{1996A&AS..117..393B},
          GALFIT \citep{2002AJ....124..266P},
          EAZY \citep{Bra08},
          CASA \citep{2007ASPC..376..127M},
          MAGPHYS\citep{Cun08},
          uvmultifit\citep{2014A&A...563A.136M}
          }

%% Appendix material should be preceded with a single \appendix command.
%% There should be a \section command for each appendix. Mark appendix
%% subsections with the same markup you use in the main body of the paper.

%% Each Appendix (indicated with \section) will be lettered A, B, C, etc.
%% The equation counter will reset when it encounters the \appendix
%% command and will number appendix equations (A1), (A2), etc. The
%% Figure and Table counter will not reset.

\bibliography{sample631}{}
\bibliographystyle{aasjournal}

%% This command is needed to show the entire author+affiliation list when
%% the collaboration and author truncation commands are used.  It has to
%% go at the end of the manuscript.
%\allauthors

%% Include this line if you are using the \added, \replaced, \deleted
%% commands to see a summary list of all changes at the end of the article.
%\listofchanges

\end{document}